\documentclass[preprint,showpacs,superscriptaddress]{revtex4}
\usepackage[utf8]{inputenc}
\usepackage{array}
\usepackage{hhline}
\usepackage{multirow}
\usepackage{booktabs}
\usepackage{float}
\usepackage{graphicx}				
\usepackage{amsfonts}
\usepackage{amssymb}
\usepackage{physics}
\usepackage{amsmath}
\usepackage[dvips]{epsfig}
\usepackage[footnotesize]{subfigure}
\usepackage[font=footnotesize]{caption}
\numberwithin{equation}{section}
\usepackage{mathrsfs}
\usepackage{bm}
\usepackage[title]{appendix}
\makeatletter
\newcommand{\oset}[2]{%
  {\mathop{#2}\limits^{\vbox to -.5\ex@{\kern-\tw@\ex@
  \hbox{\scriptsize #1}\vss}}}}
\makeatother
\makeatletter
\newsavebox\myboxA
\newsavebox\myboxB
\newlength\mylenA
\newcommand*\xoverline[2][0.75]{%
    \sbox{\myboxA}{$\m@th#2$}%
    \setbox\myboxB\null
    \ht\myboxB=\ht\myboxA%
    \dp\myboxB=\dp\myboxA%
    \wd\myboxB=#1\wd\myboxA
    \sbox\myboxB{$\m@th\overline{\copy\myboxB}$}
    \setlength\mylenA{\the\wd\myboxA}
    \addtolength\mylenA{-\the\wd\myboxB}%
    \ifdim\wd\myboxB<\wd\myboxA%
       \rlap{\hskip 0.5\mylenA\usebox\myboxB}{\usebox\myboxA}%
    \else
        \hskip -0.5\mylenA\rlap{\usebox\myboxA}{\hskip 0.5\mylenA\usebox\myboxB}%
    \fi}
\makeatother

\def\la{\lower4pt\hbox{$\buildrel\textstyle\ <\over\sim\ $}}

\def\simgreat{\buildrel \textstyle \lower3pt\hbox{$\sim$} \over >}
\def\simless{\buildrel \textstyle \lower3pt\hbox{$\sim$} \over <}
\newcommand{\be}{\begin{equation}}
\newcommand{\ee}{\end{equation}}
\newcommand{\beq}{\begin{eqnarray}}
\newcommand{\eeq}{\end{eqnarray}}
\newcommand\bsdot{\ensuremath{\boldsymbol{.}}}
\def\cchit{\tilde{\raise2pt\hbox{$\chi$}}} 
\def\cchitt{\tilde{\tilde{\raise2pt\hbox{$\chi$}}}} 
\def\gapf{\lower4pt\hbox{$\ \buildrel \textstyle> \over \sim \ $}}
\def\gapt{\lower2pt\hbox{$\buildrel \textstyle> \over \sim$}}

\def\lapf{\lower4pt\hbox{$\ \buildrel \textstyle< \over \sim \ $}}
\renewcommand{\vec}[1]{\boldsymbol{\mathbf{#1}}}
\newcommand{\unv}[1]{\hat{\vec{#1}}}

\def\cchi{\raise2pt\hbox{$\chi$}} 

\begin{document}
%
%
\renewcommand{\thesection}{\arabic{section}} 
\renewcommand{\thesubsection}{\thesection.\arabic{subsection}}
\title{Reflection and transmission of electromagnetic pulses at a planar dielectric interface -- theory and
quantum lattice simulations}

%
%
\author{Abhay K. Ram}
\affiliation{Plasma Science and Fusion Center, 
Massachusetts Institute of Technology, Cambridge, MA 02139.}
\author{George Vahala}
\affiliation{Department of Physics, 
College of William \& Mary, Williamsburg, VA23185}
\author{Linda Vahala}
\affiliation{Department of Electrical \& Computer Engineering, 
Old Dominion University, Norfolk, VA 12319}
\author{Min Soe}
\affiliation{Department of Mathematics and Physical Sciences, 
Rogers State University, Claremore, OK 74017}

%
%
\begin{abstract}
There is considerable interest in the application of quantum information science to advance computations in plasma physics.  
A particular point of curiosity is whether it is possible to take advantage of quantum computers to speed up numerical simulations 
relative to conventional computers. Many of the topics in fusion plasma physics are classical in nature. In order to implement
them on quantum computers it will require couching a classical problem in the language of quantum mechanics. 
Electromagnetic waves are routinely used in fusion experiments to heat a plasma or to generate currents in the plasma. The propagation
of electromagnetic waves is described by Maxwell equations with an appropriate description of the
plasma as a dielectric medium. Before advancing to the tensor dielectric of a magnetized plasma, this paper considers electromagnetic
wave propagation in a one-dimensional inhomogeneous scalar dielectric. 

The classic theory of scattering of plane electromagnetic waves at a planar interface, separating two different dielectric
media, leads to Fresnel equations for reflection and transmission coefficients. 
In contrast to plane waves, this paper is on the reflection and transmission of a spatially confined 
electromagnetic pulse. Following an analytical formulation for the scattering of a Gaussian pulse, it is deduced that the maximum 
transmission coefficient for a pulse is $\sqrt{n_2/n_1}$ times that for a plane wave; the incident and transmitted
pulses propagating in dielectric media with refractive indices $n_1$ and $n_2$, respectively. 
The analytical theory is complemented by numerical simulations using a quantum lattice algorithm for Maxwell equations.
The algorithm, based on the Riemann-Silberstein-Weber representation of the
electromagnetic fields and expressed in term of qubits, is an interleaved sequence of entangling
operators at each lattice site and unitary streaming operators which transmit information from one site to an adjacent lattice site. Besides
substantiating results from the theory for Gaussian pulses, numerical simulations show their validity for non-Gaussian pulses. 
Apart from their time-asymptotic forms, the simulations display an interplay between the incident, reflected, and 
transmitted pulses in the vicinity of the transition region between two dielectric media.

\end{abstract}


\maketitle

\section{Introduction}
\label{sec:1}

The anticipated availability of quantum computers in the near future, and the associated speed up in computations, has provided
an impetus for exploring applications of quantum information science to plasma physics. Our motivation in this paper is to
study the propagation of electromagnetic waves in inhomogeneous dielectric media within the realm of quantum information science.
The propagation of waves is described by Maxwell equations in which information about a dielectric medium is expressed through
its permittivity. In ordinary materials the permittivity is usually a scalar, but in a magnetized plasma it is a tensor.
This entire description of waves is classical in the sense that no quantum effects come into play. However, it was 
recognized early on, in 1931, by Oppenheimer \cite{oppie} that it is possible to cast Maxwell equations for electromagnetic fields 
in vacuum into a form that is similar to the Dirac equation \cite{good}. More recently, this formalism has been extended 
to waves propagating in a homogeneous dielectric medium and, subsequently, to waves propagating in a
spatially inhomogeneous dielectric medium \cite{khan}; in both cases, the permittivity of the medium being a scalar.

In this paper, we study the propagation of electromagnetic waves in spatially inhomogeneous scalar dielectrics. This is a prelude
to future studies on wave propagation in magnetized plasmas. We construct a theoretical model for the reflection and transmission
of a spatially confined electromagnetic pulse incident on a surface separating two different dielectric media. The plane wave version
of this model is a standard example in textbooks on electromagnetism \cite{griffiths}. For a simulation code that will complement
the theory, we cast Maxwell equations
in a matrix representation that is akin to the Dirac equation \cite{khan}. We formulate a quantum lattice algorithm based on this
representation. The results from simulations using this algorithm motivated the analytical formalism, and are found to 
be in excellent agreement with those given by the theory. A quantum lattice algorithm (QLA) is an interleaved sequence of unitary 
collision and streaming operators 
that can be modeled by qubit gates. Such an algorithm is ideally parallelized on traditional computers and can be implemented on a quantum computer.
Thus, a QLA can be tested before quantum computers become readily available.

We outline the plane wave model of reflection and transmission at an interface separating two different dielectrics. This will set up the
theory for propagation of an electromagnetic pulse.
In the Cartesian coordinate system, consider a plane interface at $z=0$ which divides space into two non-magnetic regions with different scalar dielectric
permittivities, 
\begin{equation}
\epsilon (z) \ = \ \begin{cases} \epsilon_1 & \quad {\rm for} \  z < 0 \ \ \left({\rm region}\ 1 \right) \\
\epsilon_2 & \quad {\rm for} \  z > 0 \ \ \left({\rm region}\ 2 \right)
\end{cases}
\label{1.1}
\end{equation}
An incident plane wave in region 1, propagating towards the interface along the $z$-direction -- normal to the interface -- 
will lead to a reflected and  
a transmitted plane waves. The electric and magnetic fields of the
three waves, respectively, are of the form,
\begin{align}
\mathbf{E}_I \ & =  \ E_0 \; e^{i \left( k_1 z - \omega t \right)} \ \unv{x},  \ \ & \mathbf{B}_I \ &  =  \ \frac{1}{c_1} \; 
E_0 \; e^{i \left( k_1 z - \omega t \right)} \ \unv{y}, \label{1.2a} \\
\mathbf{E}_R \ & =  \ E_r\;  e^{i \left( - k_1 z - \omega t \right)} \ \unv{x},  \ \ & \mathbf{B}_R \ &  =  \ -\frac{1}{c_1} \; 
E_r \; e^{i \left( - k_1 z - \omega t \right)} \ \unv{y}, \label{1.2b} \\
\mathbf{E}_T \ & =  \ E_t\;  e^{i \left( k_2 z - \omega t \right)} \ \unv{x},  \ \ & \mathbf{B}_T \ &  =  \ \frac{1}{c_2} \; 
E_t \; e^{i \left( k_2 z - \omega t \right)} \ \unv{y},\label{1.2c}
\end{align}
where $E_0$, $E_r$, and $E_t$ are complex field amplitudes, and $\omega$ is the angular frequency of the wave. Within each
dielectric region, the speed of light is $v_i = 1/\sqrt{\mu_o \epsilon_i}$, the refractive index is $n_i = \sqrt{ \epsilon_i / \epsilon_0 } = c/v_i$,
$k_i = \omega n_i /c$, $\mu_0$ is the vacuum permeability, $\epsilon_0$ is the vacuum permittivity,
$c$ is the speed of light in vacuum, and $i=1, \; 2$. The boundary conditions,
that follow from  Ampere and Faraday equations, require continuity of the tangential electric and magnetic fields at $z=0$, resulting in
the Fresnel relations \cite{griffiths},
\begin{alignat}{2}
E_r \ & = \ - \left( \frac{v_1 - v_2}{v_1 + v_2} \right)  E_0  & \ = \ \left( \frac{n_1 - n_2}{n_1 + n_2} \right)  E_0, \label{1.3a} \\
E_t \ & = \ \phantom{-} \left( \frac{2 v_2}{v_1 + v_2} \right)  E_0  & \ = \ \left( \frac{2 n_1}{n_1 + n_2} \right)  E_0.
\label{1.3b}
\end{alignat}

In this paper, we develop a model for the reflection and transmission of a Gaussian pulse at
the interface given in \eqref{1.1}. The incident Gaussian pulse is propagating along the normal to the interface.
We show that the amplitude of the transmitted pulse is different from that obtained for
plane waves -- it is larger by a factor $\sqrt{n_2 / n_1}$. 
The analytical model is complemented by QLA simulations for electromagnetic wave propagation
in a dielectric medium. While the analytical theory is based on satisfying electromagnetic boundary conditions at the discontinuous 
interface, the QLA simulations propagate the pulse through a continuous, monotonic, narrow interface representing 
the discontinuity. The two approaches give identical results. Moreover, the simulations display the interplay between
the incident, reflected, and transmitted pulses in the vicinity of the interface. This is not possible within the
realm of our analytical theory. 

The rest of the paper is organized as follows. The analytical model is developed in section \ref{sec:2} followed by a matrix formulation of
Maxwell equation for an inhomogeneous scalar dielectric in section \ref{sec:3}. Based on this formulation, the QLA is set up in section \ref{sec:4}.
In section \ref{sec:5} we display results from QLA simulations and compare them with the analytical model.

\section{Reflection and transmission of a Gaussian pulse}
\label{sec:2}

In what follows, we assume that the Gaussian pulse has a compact support. However, for mathematical convenience, any integrals involving
the pulse will be extended from $-\infty$ to $\infty$. Also, the incident, reflected, and transmitted pulses will be considered as separate
entities, thereby avoiding those times when the pulses overlap near the interface.

\subsection{The incident Gaussian pulse}
\label{sec:2A}

At time $t=0$, we assume that the incident pulse has the form,
\begin{equation}
{\mathbf E}_I(z) \ = \ E_0\; \alpha \ e^{-\left( z+z_0 \right)^2/\left( 2 \sigma^2 \right)} \ \unv{x},
\label{2.1}
\end{equation}
where $\alpha$ is a normalization constant, $z_0 > 0$ is the location of the peak of the pulse, $\sigma$ is its effective width, and
the pulse is assumed to be localized in the region $z < 0$. The requirement,
\begin{equation}
\int_{-\infty}^{\infty} \ dz \ {\mathbf E}_I(z) \bsdot {\mathbf E}_I^* (z) = \left| E_0 \right|^2,
\label{2.2}
\end{equation} 
leads to,
\begin{equation}
\alpha \ = \ \frac{1}{\pi^{1/4} \; \sqrt{\sigma}}. 
\label{2.3}
\end{equation}
In \eqref{2.2},  $^*$ denotes complex conjugate of the field. The time evolution of the normalized Gaussian pulse in medium 1 is,
\begin{equation}
{\mathbf E}_I(z,t) \ = \ \frac{E_0}{\pi^{1/4} \; \sqrt{\sigma}} \ e^{-\left( z-v_1 t+z_0 \right)^2/\left( 2 \sigma^2 \right)} \ \unv{x}, \ \quad {\rm for} \ z < 0.
\label{2.5}
\end{equation}
Upon taking the Fourier transform of \eqref{2.5}, the plane wave representation of the incident Gaussian pulse is, 
\begin{equation}
{\mathbf E}_{Ik}(k,z,t) \ = \ \frac{E_0}{\pi^{1/4}} \ \sqrt{2 \pi \sigma} \ e^{- \sigma^2 k^2 /2} \  e^{ik \left(z+z_0-v_1t \right)} \ \unv{x} \ \equiv \ 
E_{Ik}(k,z,t) \ \unv{x},
\label{2.6}
\end{equation}
with,
\begin{equation}
{\mathbf E}_I(z,t) \ = \ \frac{1}{2\pi} \; \int_{-\infty}^{\infty} dk \ {\mathbf E}_{Ik} \left( k,z,t \right) 
\label{2.7}
\end{equation}
The magnetic field associated with each plane wave is, 
\begin{equation}
{\mathbf B}_{Ik}(k,z,t) \ = \ \frac{1}{v_1} \ E_{Ik}(k,z,t) \ \unv{y},
\label{2.8}
\end{equation}

\subsection{The reflected and transmitted Gaussian pulses}
\label{sec:2B}

The reflected (medium 1) and the transmitted (medium 2) pulses are taken to be of the following forms, respectively,
\begin{align}
{\mathbf E}_R(z,t) \ = \ & \frac{E_r}{\pi^{1/4} \sqrt{\sigma_1}} \ e^{-\left(z +z_1+v_1t \right)^2/ \left( 2 \sigma_1^2 \right)} \ \unv{x},
\ \quad {\rm for} \ z<0, 
\label{2b1} \\
{\mathbf E}_T(z,t) \ = \ & \frac{E_t}{\pi^{1/4} \sqrt{\sigma_2}} \ e^{-\left(z -z_2-v_2t \right)^2/ \left(2 \sigma_2^2 \right)} \ \unv{x}, \ \quad {\rm for} \ z>0, 
\label{2b2}
\end{align}
where $z_1$ and $z_2$ are constants that shift the Gaussian pulses away
from $z = 0$, and $\sigma_1$ and $\sigma_2$ are effective widths of each of the pulses.
Analogus to the incident pulse, the plane wave representation of the reflected and transmitted pulses is,
\begin{align}
{\mathbf E}_{Rk}(k,z,t) \ = \ & \frac{E_r}{\pi^{1/4}} \ \sqrt{2 \pi \sigma_1} \ e^{-\sigma_1^2 k_1^2/2} 
\ e^{i k_1 \left( z +z_1 + v_1t \right)} \ \unv{x} \ \equiv \ E_{Rk}(k,z,t) \ \unv{x}, \label{2b3} \\
E_{Tk}(k,z,t) \ = \ & \frac{E_t}{\pi^{1/4}} \ \sqrt{2 \pi \sigma_2} \ e^{-\sigma_2^2 k_2^2/2} 
\ e^{i k_2 \left( z -z_2 - v_2t \right)} \ \unv{x}  \ \equiv \ E_{Tk}(k,z,t) \ \unv{x}, \label{2b4}
\end{align}
where $k_1$ and $k_2$ are the Fourier space variables for the reflected and transmitted waves, respectively.
The corresponding magnetic fields are,
\begin{align} 
{\mathbf B}_{Rk}(k,z,t) \ = \ & - \frac{1}{v_1} \ E_{Rk}(k,z,t) \ \unv{y}, \label{2b5} \\
{\mathbf B}_{Tk}(k,z,t) \ = \ & \frac{1}{v_2} \ E_{Tk}(k,z,t) \ \unv{y}. \label{2b6}
\end{align}

\subsection{Boundary conditions -- amplitudes and widths of the reflected and transmitted pulses}
\label{sec:2C}

The boundary conditions -- continuity of the tangential electric and magnetic fields -- imposed at $z=0$ have to be 
satisfied for all times. Thus,
\begin{equation}
\omega \ \equiv \ k v_1 \ = \ -k_1 v_1 \ = \ k_2 v_2. \label{2c1}
\end{equation}
Following the discussion in section \ref{sec:1}, the Fresnel jump conditions yield,
\begin{align}
\frac{E_r}{\pi^{1/4}} \ \sqrt{2 \pi \sigma_1} \ e^{-\sigma_1^2k_1^2/2} \ e^{i k_1 z_1} \ = \ & \frac{n_1-n_2}{n_1+n_2} 
\ \frac{E_0}{\pi^{1/4}} \ \sqrt{2 \pi \sigma} \ e^{-\sigma^2 k^2/2} e^{i k z_0} , \label{2c2} \\ 
\frac{E_t}{\pi^{1/4}} \ \sqrt{2 \pi \sigma_2} \ e^{-\sigma_2^2k_2^2/2} \ e^{-i k_2 z_2} \ = \ & \frac{2 n_1}{n_1+n_2} 
\ \frac{E_0}{\pi^{1/4}} \ \sqrt{2 \pi \sigma} \ e^{-\sigma^2 k^2/2} \ e^{i k z_0} . \label{2c3}
\end{align}
From Eqs. \eqref{2c2} and \eqref{2c3}, 
\begin{align}
E_r \ = \ & \frac{n_1-n_2}{n_1+n_2} \ E_0 \ \sqrt{\frac{\sigma}{\sigma_1}} \ e^{-\frac{1}{2} \left( \sigma^2 k^2 - \sigma_1^2 k_1^2 \right)}
\ e^{i \left( kz_0 - k_1z_1 \right)}, \label{2c4} \\
E_t \ = \ & \frac{2 n_1}{n_1+n_2} \ E_0 \ \sqrt{\frac{\sigma}{\sigma_2}} \ e^{-\frac{1}{2} \left( \sigma^2 k^2 - \sigma_2^2 k_2^2 \right)}
\ e^{i \left( kz_0 + k_2z_2 \right)}. \label{2c5}
\end{align}

In defining the wave pulses \eqref{2.5}, \eqref{2b1}, and \eqref{2b2}, it was assumed that the amplitudes 
$E_0$, $E_r$, and $E_t$ are constants independent of space and time.
Consequently, in \eqref{2c4} and \eqref{2c5}, $E_r$ and $E_t$ cannot depend
on the Fourier variables $k$, $k_1$, and $k_2$. Hence,
\begin{align}
\sigma^2 k^2 - \sigma_1^2 k_1^2 \ = \ & 0, \label{2c6} \\ 
\sigma^2 k^2 - \sigma_2^2 k_2^2 \ = \ & 0, \label{2c7} \\ 
kz_0 - k_1 z_1 \ = \ & 0, \label{2c8} \\ 
kz_0 + k_2 z_2 \ = \ & 0. \label{2c9}
\end{align}
From \eqref{2c1}, $k_1 \, = \, -k$ and $k_2 \, = \, k \left(v_1 / v_2 \right)$. Making use of these
equalities in \eqref{2c6} and \eqref{2c7}, respectively, leads to,
\begin{align}
\sigma_1 \ &= \ \left| \frac{k}{k_1} \right|\; \sigma\ = \ \sigma \label{2c10} \\
\sigma_2 \ &= \ \left| \frac{k}{k_2} \right| \; \sigma \ = \ \frac{v_2}{v_1} \; \sigma \ = \ \frac{n_1}{n_2} \; \sigma. \label{2c11}
\end{align}
The width of the reflected pulse is the same as that of the incident pulse.
For $n_1 > n_2$, the width of the transmitted pulse is broader than that of the incident pulse, while for
$n_2 > n_1$ the transmitted pulse is narrower. There is an intuitive explanation of this result.
Let $T$ be the time interval between the leading edge and the trailing edge of the incident pulse arriving at $z=0$.
The ``effective width'' of the incident pulse is $T/v_1$. The reflected and transmitted pulses are formed during the
time interval $T$. The effective width of the reflected pulse is $T/v_1$ which is the same as that of the incident pulse.
However, the effective width of the transmitted pulse is $T/v_2$ -- in agreement with \eqref{2c11}

Equations \eqref{2c8} and \eqref{2c9} give a relationship between the shifts,
\begin{equation}
z_1 \ = \ - z_0 \ \ {\rm and} \ \ z_2 \ = \ -\frac{v_2}{v_1} \ z_0 \ = \ -\frac{n_1}{n_2} \ z_0.
\end{equation}
Finally, Eqs. \eqref{2c4} and \eqref{2c5} lead to the following conclusion,
\begin{align}
E_r \ = \ & \frac{n_1 - n_2}{n_1 + n_2} \ E_0, \label{2c12}\\
E_t \ = \ & \frac{2 n_1}{n_1 + n_2} \ \sqrt{\frac{n_2}{n_1}} \ E_0. \label{2c13}
\end{align}
The ratio of the amplitude of the reflected pulse to that of the incident pulse is the
same as in \eqref{1.3a} for plane wave scattering. However, the ratio of the amplitude of the transmitted pulse to
that of the initial pulse is different from \eqref{1.3b}. It is larger for $ n_2 > n_1 $
and smaller for $n_2 < n_1$ by the square-root of the ratio $n_2 / n_1$. Note that
\eqref{2c13} is unchanged when $n_1$ and $n_2$ are interchanged.

\section{Maxwell equations -- representation in terms of Riemann-Silberstein-Weber vectors}
\label{sec:3}

For non-magnetic materials with a spatially dependent permittivity $\epsilon ({\mathbf r})$, Maxwell equations are of the form,
\begin{align}
\nabla \bsdot \left\{ \epsilon \left( {\mathbf r} \right) \; {\mathbf E} \left( {\mathbf r}, t \right) \right\} \ &= \ 0, 
& \nabla \bsdot {\mathbf B} \left( {\mathbf r}, t \right) \ &= \ 0 \label{3.1} \\
\nabla \times {\mathbf E} \left( {\mathbf r}, t \right) \ &= \ -\, \frac{\partial {\mathbf B} \left( {\mathbf r}, t \right)}
{\partial t},
& \nabla \times {\mathbf B} \left( {\mathbf r}, t \right)\ &= \ \mu_0 \epsilon \left( {\mathbf r} \right) \;
\frac{\partial {\mathbf E} \left( {\mathbf r}, t \right)} 
{\partial t}. \label{3.2}
\end{align}
Here we have assumed that there are no free charges or currents in the materials.
The Riemann-Silberstein-Weber (RSB) vectors \cite{kiesslinga,sebens} for a dielectric medium are defined as \cite{khan},
\begin{equation}
{\mathbf F}^{\pm} \left( {\mathbf r}, t \right) \ = \ \frac{1}{\sqrt{2}} \; 
\left[ \sqrt{\epsilon \left( {\mathbf r} \right)} \; {\mathbf E} \left( {\mathbf r}, t \right)
 \; \pm \; \frac{i}{\sqrt{\mu_0}} \; {\mathbf B} \left( {\mathbf r}, t \right) \right] \label{3.3}
\end{equation}
After some straight forward algebraic manipulations, and making use of Maxwell equations, we find,
\begin{align}
\nabla \bsdot {\mathbf F}^{\pm} \left( {\mathbf r}, t \right) \ &= \ 
\frac{1}{2 v \left( {\mathbf r} \right)} \nabla v \left( {\mathbf r} \right) \bsdot \left[ {\mathbf F}^+ \left( {\mathbf r}, t \right) \ + \ 
{\mathbf F}^- \left( {\mathbf r}, t \right) \right], \label{3.4} \\
i\; \frac{\partial {\mathbf F}^{\pm} \left( {\mathbf r}, t \right) }{\partial t} \ &= \ \pm v \left( {\mathbf r} \right)
\nabla \times {\mathbf F}^{\pm} \ \pm \ \frac{1}{2} \; \nabla v \left( {\mathbf r} \right) \times
\left[ {\mathbf F}^+ \left( {\mathbf r}, t \right) \; + \; {\mathbf F}^- \left( {\mathbf r}, t \right) \right], \label{3.5}
\end{align}
where $v \left( {\mathbf r} \right) = 1/\sqrt{ \mu_0 \epsilon \left( {\mathbf r} \right) }$ has the dimensions of speed. 
In this representation, Eq. \eqref{3.5} is the evolution equation with \eqref{3.4} being the constraint. 

If $\epsilon \left( {\mathbf r} \right) = \epsilon_0$, then, from \eqref{3.4}, we note that the equations 
for ${\mathbf F}^+$ and ${\mathbf F}^-$ decouple and, from \eqref{3.5},
the divergences of ${\mathbf F}^+$ and of $\mathbf{F}^-$ are zero. It is easy to see the relationship between 
Eqs. \eqref{3.4} and \eqref{3.1}, and between Eqs. \eqref{3.5} and \eqref{3.2}. The advantage of using the RSW vectors is that, in vacuum,
the two ``polarizations'' ${\mathbf F}^+$ and ${\mathbf F}^-$ propagate independently \cite{birula}. 

The evolution equations \eqref{3.5} for $\mathbf{F}^+$ and $\mathbf{F}^-$ can each be cast in a $3 \times 3$ matrix form. 
In order to include \eqref{3.4} in a unified matrix representation, we need two $4$-vectors $\Psi^+$ and $\Psi^-$ defined
as follows \cite{khan},
\begin{equation}
{\Psi}^+ \left( {\mathbf r}, t \right) = 
\begin{bmatrix}
-F_x^++ iF_y^+ \\ F_z^+ \\ F_z^+ \\ F_x^++iF_y^+
\end{bmatrix},  
\quad\quad  
{\Psi}^- \left( {\mathbf r}, t \right) = 
\begin{bmatrix}
-F_x^--iF_y^- \\ F_z^- \\ F_z^- \\ F_x^--iF_y^-
\end{bmatrix}, \label{3.6}
\end{equation} 
where ${\mathbf F}^{\pm} = F_x^{\pm} \; \unv{x} + F_y^{\pm} \; \unv{y} + F_z^{\pm} \; \unv{z}$. The evolution equations for
$\Psi^+$ and $\Psi^-$ that follow from \eqref{3.4} and \eqref{3.5} are of the form \cite{khan},
\begin{align}
\frac{\partial}{\partial t} \; 
&
\begin{bmatrix}
\begin{array}{rr}
\ {\mathbf I}_4 \ & \ {\mathbf 0}_4 \ \\
{\mathbf 0}_4 & {\mathbf I}_4 \\
\end{array}
\end{bmatrix}
\; 
\begin{bmatrix}
{\Psi}^+ \\
{\Psi}^-
\end{bmatrix}
\ = \ \nonumber \\
& -v \left( {\mathbf r} \right)\; 
\begin{bmatrix}
{\mathbf M}\; \bsdot \; \nabla + {\bm \Sigma} \; \bsdot \; \frac{1}{2} \nabla \left\{ \ln v \left( {\mathbf r} \right) \right\}  & 
\ \ \ -i M_z \left[ {\bm \Sigma} \; \bsdot \; \frac{1}{2} \nabla \left\{ \ln v \left( {\mathbf r} \right) \right\} \right] \alpha_y\\
-i M_z \left[ {\bm \Sigma}^* \; \bsdot \; \frac{1}{2} \nabla \left\{ \ln v \left( {\mathbf r} \right) \right\} \right] \alpha_y & 
\ \ \ {\mathbf M}^*\; \bsdot \; \nabla + {\bm \Sigma}^* \; \bsdot \; \frac{1}{2} \nabla \left\{ \ln v \left( {\mathbf r} \right) \right\}
\end{bmatrix}
\; 
\begin{bmatrix}
{\Psi}^+ \\
{\Psi}^-
\end{bmatrix}
. \label{3.7}
\end{align}
In this equation, ${\mathbf I}_4$ is a $4 \times 4$ identity matrix, ${\mathbf{0}}_4$ is a $4 \times 4$ null matrix and,
\begin{equation}
{\bm \Sigma} \ = \
\begin{bmatrix}
\begin{array}{rr}
\ {\bm \sigma} \ & \ {\mathbf 0}_2 \ \\
  {\mathbf 0}_2 & {\bm \sigma} \\
\end{array}
\end{bmatrix}, \quad
{\bm \alpha} \ = \
\begin{bmatrix}
\begin{array}{rr}
\ {\mathbf 0}_2 \ & \ {\bm \sigma} \ \\
 {\bm \sigma} & \ \ {\mathbf 0}_2 \\
\end{array}
\end{bmatrix}
, \label{3.8}
\end{equation}
${\bm \sigma} \ = \ \sigma_x \; \unv{x} +\sigma_y \; \unv{y} +\sigma_z \; \unv{z}$ are the Pauli matrices,
\begin{equation}
\sigma_x \ = \
\begin{bmatrix}
\begin{array}{cc}
\ 0 \ & \ 1 \ \\
 1 & 0 \\
\end{array}
\end{bmatrix}, \quad
\sigma_y \ = \
\begin{bmatrix}
\begin{array}{rr}
\ 0 \ &  \ -i \ \\
 i &  0 \\
\end{array} 
\end{bmatrix}, \quad
\sigma_z \ = \
\begin{bmatrix}
\begin{array}{rr}
\ 1 \ & \ 0 \ \\
 0 & -1 \\
\end{array}  
\end{bmatrix}.
\label{3.9}
\end{equation}
In the expressions for ${\bm \Sigma}$ and ${\bm \sigma}$, $\mathbf{0}_2$ is a $2 \times 2$ null matrix.
The three Cartesian components of the matrix ${\mathbf M}$ are,
\begin{equation}
M_x \ = \ 
\begin{bmatrix} 
\begin{array}{cccc}
\ 0 \ & \ 0 \ & \ 1 \ & \ 0 \ \\ 
 0 & 0 & 0 & 1 \\ 
 1 & 0 & 0 & 0 \\ 
 0 & 1 & 0 & 0 \\
\end{array}
\end{bmatrix}, \ 
M_y \ = \ 
\begin{bmatrix}
\begin{array}{rrrr}
\ 0 \ & \ 0 \ & -i \ & 0 \ \\
 0 & 0 & 0 & -i \\
 i & 0 & 0 & 0  \\
 0 & i & 0 & 0 \\  
\end{array}
\end{bmatrix}, \
M_z \ = \ 
\begin{bmatrix}
\begin{array}{rrrr}
\ 1 \ & \ 0 \ & 0 & 0 \ \\
 0 & 1 & 0 & 0   \\
 0 & 0 & -1 & 0  \\
 0 & 0 & 0 & -1   \\
\end{array}
\end{bmatrix}.
\label{3.10}
\end{equation}

By the taking the difference of first and fourth (fifth and eighth) rows of \eqref{3.7} we obtain the evolution equation for $F_x^+$
($F_x^-$) in \eqref{3.5}; the sum of first and fourth (fifth and eighth) rows gives the evolution of $F_y^+$ ($F_y^-$).
The sum of second and third (sixth and seventh) rows leads to the evolution equation for $F_z^+$ ($F_z^-$) in \eqref{3.5}. The
difference between the second and third (sixth and seventh) rows leads to the divergence equation \eqref{3.4}. Thus, the compact
form \eqref{3.7} accounts for all the four Maxwell equations.

Equation \eqref{3.7} can be separated into three equations corresponding to the principal directions of the Cartesian coordinate system. 
For a medium in which the permittivity varies along one particular direction, the evolution equation \eqref{3.7} is simplified. However,
the three principal directions do not lead to the same evolution equation since the three Pauli matrices are different \cite{vahala-jpp,vahala-reds}. 

\subsection{Propagation and inhomogeneity along the $z$-direction}
\label{sec:3a}

As in section \ref{sec:2}, we consider one-dimensional (1-D) propagation of electromagnetic waves along the $z$-direction 
in a medium with permittivity that is a function of $z$ only.
Then Eq. \eqref{3.7} reduces to,
\begin{align}
\frac{\partial}{\partial t} \; 
&
\begin{bmatrix}
{\Psi}^+ \\
{\Psi}^-
\end{bmatrix}
\ = \ \nonumber \\
&  - v(z) \; \frac{\partial}{\partial z} \; 
\begin{bmatrix}
M_z & 
{\mathbf 0}_4 \\
{\mathbf 0}_4 &
M_z^* 
\end{bmatrix}\;
\begin{bmatrix}
{\Psi}^+ \\
{\Psi}^-
\end{bmatrix} \ - \ \frac{1}{2} \frac{\partial v(z)}{\partial z} \;
\begin{bmatrix}
\Sigma_z & 
 -i M_z \Sigma_z \alpha_y  \\
 -i M_z  \Sigma_z^*  \alpha_y  & 
\Sigma_z^* 
\end{bmatrix} \;
\begin{bmatrix}
{\Psi}^+ \\
{\Psi}^-
\end{bmatrix}. \label{3a.1}
\end{align}

If we define the four elements
of $\Psi^+$ and $\Psi^-$ as follows,
\begin{equation}
{\Psi}^+ \left( z, t \right) = 
\begin{bmatrix}
\psi_0 \left( z, t \right)\\ \psi_1 \left( z, t \right) \\ \psi_2 \left( z, t \right) \\ \psi_3 \left( z, t \right) 
\end{bmatrix}, 
\quad\quad 
{\Psi}^- \left( z, t \right) = 
\begin{bmatrix}
\psi_4 \left( z, t \right) \\ \psi_5 \left( z, t \right) \\ \psi_6 \left( z, t \right) \\ \psi_7 \left( z, t \right)
\end{bmatrix}, \label{3a.2}
\end{equation}
then \eqref{3a.1} leads to,
\begin{align}
\frac{\partial}{\partial t} \; 
\begin{bmatrix}
\begin{array}{c}
\ \psi_0 \ \\ \psi_1 \\ \psi_2 \\ \psi_3  \\
\end{array}
\end{bmatrix}, 
 \ &= \ - \; v(z) \; \frac{\partial}{\partial z} \;
\begin{bmatrix}
\begin{array}{r}
\psi_0 \\
\psi_1 \\
- \psi_2 \\
- \psi_3 \\
\end{array}
\end{bmatrix} 
\ - \ \frac{1}{2} \frac{\partial v(z)}{\partial z}  \
\begin{bmatrix}
\begin{array}{r}
\psi_0 - \psi_7 \\
-\psi_1 - \psi_6 \\
\psi_2 + \psi_5 \\
-\psi_3 + \psi_4 \\
\end{array}
\end{bmatrix}
\label{3a.3}
\\
& \nonumber \\
\frac{\partial}{\partial t} \; \begin{bmatrix}
\psi_4 \\ \psi_5 \\ \psi_6 \\ \psi_7 
\end{bmatrix}
 \ &= \ - \; v(z) \; \frac{\partial}{\partial z} \;
\begin{bmatrix}
\begin{array}{r}
\psi_4 \\
\psi_5 \\
 - \psi_6 \\
 - \psi_7 \\
\end{array}
\end{bmatrix}
\ - \ \frac{1}{2} \frac{\partial v(z)}{\partial z} \
\begin{bmatrix}
\begin{array}{r}
\psi_4 - \psi_3 \\
-\psi_5 - \psi_2 \\
\psi_6 + \psi_1 \\
-\psi_7 + \psi_0 \\
\end{array}
\end{bmatrix}
\label{3a.4}
\end{align}
In Eqns. \eqref{3a.3} and \eqref{3a.4}, it is worth noting that the time derivative of each component $\psi_i$ ($i = 0 \dots 7$) is related to
the spatial derivative of the same component. Instead, if we had assumed spatial variation in the $x$ or $y$ directions, the time derivative of one
component $\psi_i$ would be related to the spatial derivative of a different component $\psi_j$ ($j \ne i$) \cite{vahala-jpp,vahala-reds}. 
This is a consequence of $\sigma_z$ being a diagonal matrix while $\sigma_x$ and $\sigma_y$ have only off-diagonal, non-zero, elements. 

In the sections to follow, we will assume that the speed of light in the medium $v(z)$ is normalized to $c$. Even though we will continue to
use $z$ and $t$ as the space-time variables, it will be implicit that the dimensions of $z$ and $t$ are the same. The time will be related
to the discrete temporal step used in the evolution equations for the fields.
In this system of units, $v(z) = 1 / n(z)$ with $n(z)$ being
the ``spatially'' varying index of refraction.

\section{Quantum lattice algorithm for Maxwell equations}
\label{sec:4}

In setting up the quantum lattice algorithm (QLA) for wave propagation in the $z$-direction, the four-spinor
representations in Eqns. \eqref{3a.3} and \eqref{3a.4} are not
in a suitable form. For wave propagation in the $x$-direction, each spinor $\psi_i$ was classified as a qubit and the spatial derivative
entangled the qubits \cite{vahala-jpp}. In order to include entanglement, we have to increase the dimensionality of the spinor representation
from 8 to 16. This follows an earlier precedence where the Gross-Pitaevskii equation for Bose-Einstein condensates \cite{vahala-bec} was formulated
at the mesoscopic level by twice as many qubits as field components. At time $t = 0$, when the electromagnetic fields of an initial incoming
pulse are set up, we initialize the qubits in the 16 spinor representation to be,
\begin{equation}
\begin{aligned}
q_0 \left(z, 0 \right) &= \dfrac{1}{2} \psi_0 \left( z, 0 \right), & q_1 \left(z, 0 \right) &= \dfrac{1}{2} \psi_1 \left(z, 0 \right), 
& q_2 \left(z, 0 \right) &= q_0 \left(z, 0 \right), & q_3 \left(z, 0 \right) &= q_1 \left(z, 0 \right), \\
q_4 \left(z, 0 \right)  &=  \dfrac{1}{2} \ \psi_2 \left(z, 0 \right), & q_5 \left(z, 0 \right) &= \dfrac{1}{2} \psi_3 \left(z, 0 \right), 
& q_6 \left(z, 0 \right) &=  q_4 \left(z, 0 \right), & q_7 \left(z, 0 \right) &= q_5 \left(z, 0 \right), \\
q_8 \left(z, 0 \right) &= \dfrac{1}{2} \psi_4 \left(z, 0 \right), & q_9 \left(z, 0 \right) &= \frac{1}{2} \psi_5 \left(z, 0 \right), 
& q_{10} \left(z, 0 \right) &= q_8 \left(z, 0 \right), & q_{11} \left(z, 0 \right) &=  q_9 \left(z, 0 \right), \\
q_{12} \left(z, 0 \right) &= \dfrac{1}{2} \psi_6 \left(z, 0 \right), & q_{13} \left(z, 0 \right) &= \frac{1}{2} \psi_7 \left(z, 0 \right), 
& q_{14} \left(z, 0 \right) &= q_{12} \left(z, 0 \right), & q_{15} \left(z, 0 \right) &= q_{13} \left(z, 0 \right). 
\end{aligned}
\label{4.1}
\end{equation}
The choice expressed in \eqref{4.1} requires the collision matrix at every spatial lattice point to couple each pair of the qubits:
$\left(q_0,\, q_2 \right)$, $\left(q_1,\, q_3 \right)$, $\left(q_4,\, q_6 \right)$, $\left(q_5,\, q_7 \right)$,
$\left(q_8,\, q_{10} \right)$, $\left(q_{9},\, q_{11} \right)$, $\left(q_{12},\, q_{14} \right)$, and $\left(q_{13},\, q_{15} \right)$.
An appropriate form of the unitary collision matrix that couples the qubits is,
\begin{equation}
{\mathbf C} \left( \theta \right) \ = \ 
\begin{bmatrix}
{\mathbf V}_4 \left( \theta \right) & {\mathbf 0}_4&  {\mathbf 0}_4 & {\mathbf 0}_4  \\
{\mathbf 0}_4  &  {\mathbf V}_4^T \left( \theta \right)  & {\mathbf 0}_4 & {\mathbf 0}_4  \\  
{\mathbf 0}_4 & {\mathbf 0}_4 & {\mathbf V}_4 \left( \theta \right)  &  {\mathbf 0}_4 \\
{\mathbf 0}_4 & {\mathbf 0}_4 & {\mathbf 0}_4 & {\mathbf V}_4^T \left( \theta \right) \\ 
\end{bmatrix},
\label{4.2}
\end{equation} 
where,
\begin{equation}
{\mathbf V}_4 \left( \theta \right) \ = \ 
\cos \left( \theta \right) \; 
\begin{bmatrix}
\begin{array}{rrrr}
1 & \ 0 & \ 0 & \ 0 \\
0 & 1 & 0 & 0 \\  
1 & 0 & 1 & 0 \\
0 & 0 & 0 & 1 \\ 
\end{array}
\end{bmatrix} \; + \; 
\sin \left( \theta \right) \; 
\begin{bmatrix}
\begin{array}{rrrr}
0 & 0 & \ 1 & \ 0 \\
0 & 0 & 0 & 1 \\  
-1 & 0 & 0 & 0 \\
0 & -1 & 0 & 1 \\ 
\end{array}
\end{bmatrix}, 
\label{4.3}
\end{equation} 
${\mathbf V}_4^T \left( \theta \right)$ is the transpose of ${\mathbf V}_4 \left( \theta \right)$, and
the mixing angle $\theta$ will be given later in this section.

There are two different streaming operators which translate qubits from one lattice site to a neighboring site. Each streaming operator is unitary
and diagonal, and operates only on one element of each pair of qubits discussed above. The streaming operators ${\mathbf S}^{(1),(2)}_{\pm \epsilon}$
acting on the 16 qubit representation lead to,
\begin{equation}
{\mathbf S}^{(1)}_{\pm \epsilon} \ 
\begin{bmatrix}
\begin{array}{l} 
q_0 \left( z, t \right)\\ 
q_1 \left( z, t \right)\\ 
q_2 \left( z, t \right) \\ 
q_3 \left( z, t \right) \\ 
q_4 \left( z, t \right) \\ 
q_5 \left( z, t \right) \\ 
q_6 \left( z, t \right) \\ 
q_7 \left( z, t \right) \\ 
q_8 \left( z, t \right) \\ 
q_9 \left( z, t \right) \\ 
q_{10} \left( z, t \right) \\ 
q_{11} \left( z, t \right) \\ 
q_{12} \left( z, t \right) \\ 
q_{13} \left( z, t \right) \\ 
q_{14} \left( z, t \right) \\ 
q_{15} \left( z, t \right) \\
\end{array}
\end{bmatrix}
\ = \ 
\begin{bmatrix}
\begin{array}{l} 
q_0 \left( z \pm \epsilon, t \right)\\
q_1 \left( z \pm \epsilon, t \right)\\
q_2 \left( z, t \right) \\
q_3 \left( z, t \right) \\
q_4 \left( z \pm \epsilon, t \right) \\
q_5 \left( z \pm \epsilon, t \right) \\
q_6 \left( z, t \right) \\
q_7 \left( z, t \right) \\
q_8 \left( z \pm \epsilon, t \right) \\
q_9 \left( z \pm \epsilon, t \right) \\
q_{10} \left( z, t \right) \\
q_{11} \left( z, t \right) \\
q_{12} \left( z \pm \epsilon, t \right) \\
q_{13} \left( z \pm \epsilon, t \right) \\
q_{14} \left( z, t \right) \\
q_{15} \left( z, t \right) \\
\end{array}
\end{bmatrix},
\quad \quad
{\mathbf S}^{(2)}_{\pm \epsilon} \ 
\begin{bmatrix}
\begin{array}{l}
q_0 \left( z, t \right)\\
q_1 \left( z, t \right)\\
q_2 \left( z, t \right) \\
q_3 \left( z, t \right) \\
q_4 \left( z, t \right) \\
q_5 \left( z, t \right) \\
q_6 \left( z, t \right) \\
q_7 \left( z, t \right) \\
q_8 \left( z, t \right) \\
q_9 \left( z, t \right) \\
q_{10} \left( z, t \right) \\
q_{11} \left( z, t \right) \\
q_{12} \left( z, t \right) \\
q_{13} \left( z, t \right) \\
q_{14} \left( z, t \right) \\
q_{15} \left( z, t \right) \\
\end{array}
\end{bmatrix}
\ = \ 
\begin{bmatrix}
\begin{array}{l}
q_0 \left( z, t \right)\\
q_1 \left( z, t \right)\\
q_2 \left( z \pm \epsilon, t \right) \\
q_3 \left( z \pm \epsilon, t \right) \\
q_4 \left( z, t \right) \\
q_5 \left( z, t \right) \\
q_6 \left( z \pm \epsilon, t \right) \\
q_7 \left( z \pm \epsilon, t \right) \\
q_8 \left( z, t \right) \\
q_9 \left( z, t \right) \\
q_{10} \left( z \pm \epsilon, t \right) \\
q_{11} \left( z \pm \epsilon, t \right) \\
q_{12} \left( z, t \right) \\
q_{13} \left( z, t \right) \\
q_{14} \left( z \pm \epsilon, t \right) \\
q_{15} \left( z \pm \epsilon, t \right) \\
\end{array}
\end{bmatrix},
\label{4.4}
\end{equation}
where $\epsilon$ is the step-size to the adjacent lattice site. We have also used $\epsilon$ for the permitivity of a medium. 
In the rest of the narrative, $\epsilon$ is the step-size and any dielectric medium will be described by its index of refraction.

In order to include spatial inhomogeneity in the refractive index, we need two collision operators
which provide the 16 qubit coupling, similar in vein to the 8-spinor coupling in the second term on the right
hand side of Eqs. \eqref{3a.3} and \eqref{3a.4}. These operators, ${\mathbf P}^{(1)}$ and ${\mathbf P}^{(2)}$, 
referred to as potential collision operators,
are,
\begin{equation}
{\mathbf P}^{(1)} \left( \gamma \right) \ = \ 
\begin{bmatrix}
\begin{array}{cccc}
{\mathbf \Phi}_4 & {\mathbf 0}_4 & {\mathbf 0}_4 & {\mathbf 0}_4 \\ 
{\mathbf 0}_4 & {\mathbf \Phi}_4 & {\mathbf 0}_4 & {\mathbf 0}_4 \\  
{\mathbf 0}_4 & {\mathbf 0}_4 & {\mathbf \Phi}_4 & {\mathbf 0}_4 \\   
{\mathbf 0}_4 & {\mathbf 0}_4 & {\mathbf 0}_4 & {\mathbf \Phi}_4 \\   
\end{array}
\end{bmatrix}, 
\label{4.5} 
\end{equation}
\begin{equation}
{\mathbf P}^{(2)} \left( \gamma \right) \ = \ 
\begin{bmatrix}
\begin{array}{cc}
\overline{\mathbf \Phi}^{(1)}_8 \left( \gamma \right) & \overline{\mathbf \Phi}^{(2)}_8 \left( \gamma \right) \\
\overline{\mathbf \Phi}^{(2)}_8 \left( \gamma \right) & \overline{\mathbf \Phi}^{(1)}_8 \left( \gamma \right) \\
\end{array}
\end{bmatrix}
\label{4.6}
\end{equation}
where,
\begin{equation}
{\mathbf \Phi}_4 \ = \ \cos \left( \gamma \right) \; {\mathbf I}_4 \ + \ \sin \left( \gamma \right) \;
\begin{bmatrix}
\begin{array}{rrrr}
0 & 0 & -1 & 0 \\
0 & 0 & 0 & -1 \\
-1 & 0 & 0 & 0 \\ 
0 & -1 & 0 & 0 \\ 
\end{array}
\end{bmatrix},
\label{4.7}
\end{equation}
\begin{equation}
\quad \quad 
\overline{\mathbf \Phi}^{(1)} \ = \ \cos \left( \gamma \right) \;{\mathbf I}_8,  \quad \quad
\overline{\mathbf \Phi}^{(2)} \ = \ \sin \left( \gamma \right) \; 
\begin{bmatrix}
\begin{array}{llllrrrr}
0 \; \; &  0 \; \;  &  0 \; \; &  0 \; \; & 0 & 0 & 0 & -1 \\
0 & 0 & 0 & 0 & 0 & 0 & -1 & 0 \\
0 & 0 & 0 & 0 & 0 & -1 & 0 & 0 \\
0 & 0 & 0 & 0 & -1 & 0 & 0 & 0 \\
0 & 0 & 0 & 1 & 0 & 0 & 0 & 0 \\
0 & 0 & 1 & 0 & 0 & 0 & 0 & 0 \\
0 & 1 & 0 & 0 & 0 & 0 & 0 & 0 \\
1 & 0 & 0 & 0 & 0 & 0 & 0 & 0 \\
\end{array}
\end{bmatrix},
\label{4.8}
\end{equation}
where $\gamma$ is a mixing angle.

From the collide-stream operators, we can construct the following unitary operators,
\begin{equation}
\begin{aligned}
{\mathbf U} \ &= \ {\mathbf S}^{(1)}_{-\epsilon} \; {\mathbf C} \left( \theta \right) \; {\mathbf S}^{(1)}_{\epsilon} \; {\mathbf C}^{\dagger} \left( \theta \right)  
\; {\mathbf S}^{(2)}_{\epsilon} \; {\mathbf C} \left( \theta \right) \; {\mathbf S}^{(2)}_{-\epsilon} \; {\mathbf C}^{\dagger} \left( \theta \right), \\  
{\mathbf U}^{\dagger} \ &= \ {\mathbf S}^{(1)}_{\epsilon} \; {\mathbf C} ^{\dagger} \left( \theta \right) \; {\mathbf S}^{(1)}_{-\epsilon} \; 
{\mathbf C} \left( \theta \right)  \; 
{\mathbf S}^{(2)}_{-\epsilon} \; {\mathbf C}^{\dagger} \left( \theta \right) \; {\mathbf S}^{(2)}_{\epsilon} \; {\mathbf C} \left( \theta \right),
\end{aligned}
\label{4.9}
\end{equation}
where $^{\dagger}$ is the complex conjugate transpose of the matrix. The time evolution of the 16 qubits can be expressed in terms of these
unitary operators,
\begin{equation}
{\mathbf Q} \left( z,\; t + \delta t \right) \ = \ {\mathbf P}^{(2)} \; {\mathbf P}^{(1)} \; {\mathbf U}^{\dagger} \; {\mathbf U} \; 
{\mathbf Q} \left( z, \; t \right),
\label{4.10}
\end{equation}
where ${\mathbf Q}^T \left(z, \; t \right) \ = \ \left[q_0 \left(z, \; t \right),\; q_1 \left(z, \; t \right), \; \ldots \ldots \; 
q_{15} \left(z, \; t \right) \right]$.

The quantum lattice algorithm is complete once the mixing angles $\theta$ and $\gamma$ are defined. For this, we assume that $\epsilon$ is a
perturbation parameter and make the ansatz that $\theta \propto \epsilon$ and $\gamma \propto \epsilon^2$. This ordering is akin to
diffusion ordering. Subsequently, we expand the evolution equation \eqref{4.10} out to order $\epsilon^2$ using Mathematica. There is an
additional constraint that guides our choice of the mixing angles. At order $\epsilon^2$, with an appropriate combinations of the
16 qubits, we should retrieve, in the continuum limit, the 8-spinor form of Maxwell equations given in \eqref{3a.3} and \eqref{3a.4}.
We find that,
\begin{equation}
\theta \ = \ \dfrac{1}{4\; n(z)} \; \epsilon, \quad \quad \gamma \ = \ \dfrac{1}{2} \dfrac{n'(z)}{n^2(z)} \; \epsilon^2,
\label{4.11}
\end{equation}
where $'$ denotes a derivative with respect to the argument. Note that, in our normalized set of units, $n(z) = 1 / v(z)$.
In the continuum limit, 
and to order $\epsilon^2$, \eqref{4.10} leads to the following mesoscopic evolution equation for the 16 qubits,
\begin{equation}
\dfrac{\partial}{\partial t} \ 
\begin{bmatrix}
\begin{array}{l}  
q_0 \\  
q_1 \\  
q_2 \\  
q_3 \\  
q_4 \\  
q_5 \\  
q_6 \\  
q_7 \\  
q_8 \\  
q_9 \\  
q_{10} \\  
q_{11} \\  
q_{12} \\  
q_{13} \\  
q_{14} \\  
q_{15} \\
\end{array}
\end{bmatrix}
\ = \ - \dfrac{1}{n \left( z \right)} \; \dfrac{\partial}{\partial z} \;  
\begin{bmatrix}
\begin{array}{r}  
q_2 \\  
q_3 \\  
q_0 \\  
q_1 \\ 
-q_6 \\ 
-q_7 \\ 
-q_4 \\ 
-q_5 \\  
q_6 \\  
q_{11} \\  
q_{8} \\  
q_{9} \\ 
-q_{14} \\ 
-q_{11} \\ 
-q_{12} \\ 
-q_{13} \\
\end{array}
\end{bmatrix}
\ + \ \dfrac{1}{2} \dfrac{n' \left( z \right) }{n^2 \left( z \right)} \;  
\begin{bmatrix}
\begin{array}{r} 
-q_2 + q_{15} \\  
q_3 + q_{14} \\ 
-q_0 + q_{13} \\  
q_1 + q_{12} \\ 
-q_6 - q_{11} \\  
q_7 - q_{10} \\ 
-q_4 - q_9 \\  
q_5 - q_8 \\ 
-q_{10} + q_7 \\  
q_{11} + q_6 \\ 
-q_8 + q_5 \\  
q_9 + q_4 \\ 
-q_{14} - q_3 \\  
q_{15} - q_2 \\ 
-q_{12} - q_1 \\  
q_{13} - q_0 \\
\end{array}
\end{bmatrix}
\label{4.13}
\end{equation}
where all the $q_i$'s $\left( i = 0, 1, \ldots \ldots 15 \right)$ are functions of $z$ and $t$. 
It is straight forward to show that \eqref{4.13} leads to the 8-spinor evolution equations \eqref{3a.3} and \eqref{3a.4} 
with the following substitutions,
\begin{equation}
\begin{array}{llll}
\psi_0 \  = \ q_0 \ + \ q_2, & \ \ \psi_1 \ = \ q_1 \ + \ q_3, & \ \ \psi_2 \ = \ q_4 \ + \ q_6, & \ \ \psi_3 \ = \ q_5 \ + \ q_7, \\
\psi_4 \ = \ q_8 \ + \ q_{10}, & \ \ \psi_5 \ = \ q_9 \ + \ q_{11},  & \ \ \psi_6 \ = \ q_{12} \ + \ q_{14},  & \ \ \psi_7 \ = \ q_{13} \ + \ q_{15}. \\
\end{array}
\label{4.14}
\end{equation}

\section{QLA simulations for time evolution of an electromagnetic pulse}
\label{sec:5}

The analytical expressions for the amplitude and width of the reflected and transmitted pulses in section \ref{sec:2} 
are obtained by imposing electromagnetic boundary conditions at the interface. In essence, the scattering is treated
as a boundary value problem and the results for the relative amplitudes are in the time-asymptotic limit. 
In contrast, the QLA simulations treat the scattering as an initial value problem. At time $t = 0$, a well-defined
pulse is initiated in the vicinity of the boundary of a simulation domain. The pulse propagates towards the dielectric
interface following the QLA discussed in the previous section. The interface is set up to be a smooth, monotonic, narrow transition
layer. The time evolution of the pulse, as it propagates through the transition layer, 
is governed by the same QLA --  no boundary conditions are imposed anywhere inside the
simulation domain. In the ``time-asymptotic''
limit, when the reflected and transmitted pulses are well separated, we measure their 
amplitudes and compare with theory. At the edge of the simulation
domain, we impose periodic boundary conditions. However, we stop our simulations well before the pulses reach the domain boundaries.
In the QLA computations discussed below, we have set $\epsilon = 0.2$.

\begin{figure}[htp]
\begin{center}
\captionsetup[subfigure]{justification=centering}
\subfigure[\label{fig:fig1a} \ A dielectric slab embedded in vacuum. ]{
       {\includegraphics[width=0.45\textwidth]{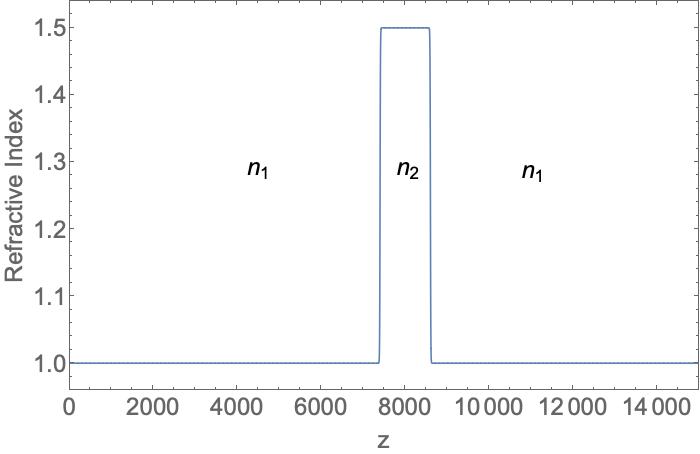}} 
} 
\subfigure[\label{fig:fig1b} \ A magnified view of the transition layer. ]{
       {\includegraphics[width=0.45\textwidth]{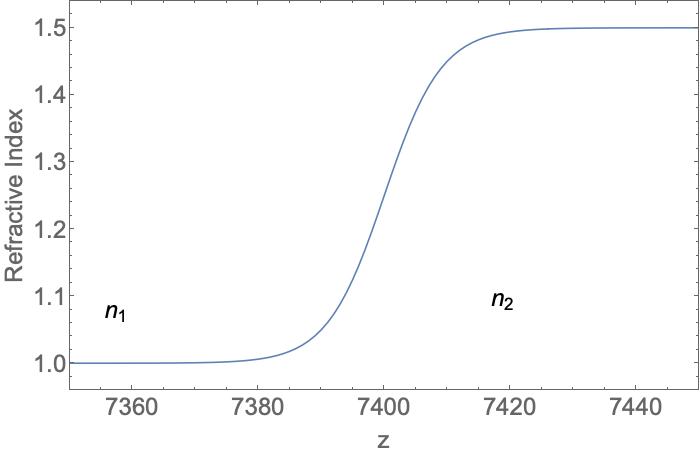}} 
} 
        \captionsetup{justification=raggedright,singlelinecheck=false} 
        \caption{ \label{fig:fig1} Permittivity as a function of $z$ with $n_1 = 1$ and $n_2 = 1.5$. 
}
\end{center}
\end{figure}

For the QLA simulations, we assume a dielectric slab with $n_2 = 1.5$ surrounded by vacuum with $n_1 = 1$. Figure \ref{fig:fig1a} shows
the index of refraction along the $z$-axis, while Fig. \ref{fig:fig1b} is a magnified view in the vicinity of the transition layer. 
In the discussion to follow, we refer to the side of the dielectric slab facing the incoming pulse as the
front-end of the slab, while the opposite side is the back-end.

We will present results from QLA simulations for two different profiles of the initial electromagnetic pulse. In the normalized system of
units, the first profile for the $x$-component of the electric field is a hyperbolic secant
function,
\begin{equation}
E_x \left( t = 0 \right) \ = \ E_0\; {\rm sech} \left\{ 0.3 \; \left( \dfrac{z - z_0}{24} \right) \right\}, \label{profile1}
\end{equation}
while the second profile is an exponential cusp of the form,
\begin{equation}
E_x \left( t = 0 \right) \ = \ E_0\; \exp \left\{ - 0.3\; \Big| \dfrac{z - z_0}{70} \Big| \right\}. \label{profile2}
\end{equation}
In vacuum, the profile for $B_y$ is the same as for $E_x$. In the simulations, we will set $E_0 = 0.01$.

We carried out a set of simulations for a Gaussian pulse, corresponding to the theory in section \ref{sec:2}. The results,
as it turns out, are the same as for the hyperbolic secant pulse. Consequently, we will concentrate our discussion on the hyperbolic secant pulse.

\subsection{Time evolution of a hyperbolic secant pulse} 
\label{sec:5A}

\begin{figure}[htp]
\begin{center}
\captionsetup[subfigure]{justification=centering}
\subfigure[\label{fig:fig2a} \ Incident pulse at $t = 6500$. ]{
       {\includegraphics[width=0.45\textwidth]{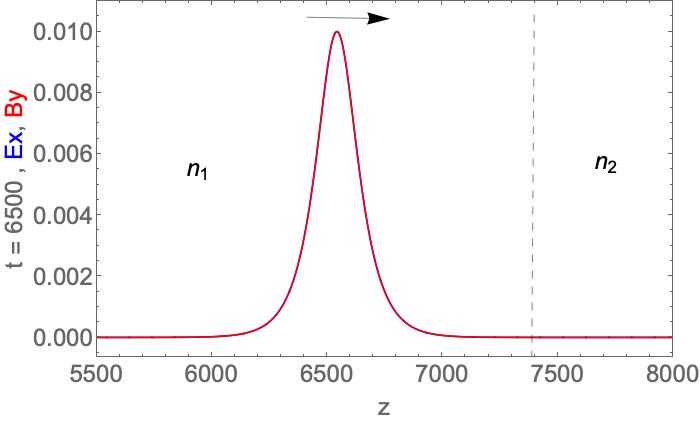}} 
} 
\subfigure[\label{fig:fig2b} \ Reflected and transmitted pulses at $t = 12000$. ]{
       {\includegraphics[width=0.45\textwidth]{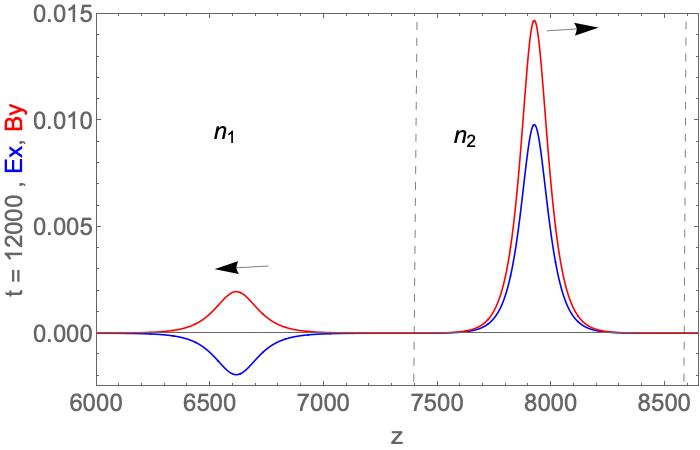}} 
} 
        \captionsetup{justification=raggedright,singlelinecheck=false} 
        \caption{ \label{fig:fig2} The electric and magnetic field components $E_x$ (blue) and $B_y$ (red), respectively, 
for (a) the incident pulse and (b) the reflected and transmitted pulses. For the incident pulse $E_x = B_y$. 
The boundaries of the dielectric slab are denoted by the vertical dashed lines. 
}
\end{center}
\end{figure}

For an initial profile of the form in \eqref{profile1}, the pulse at time $t = 6500$ is plotted in 
Fig. \ref{fig:fig2a}, as it approaches the front-end of the dielectric slab. The $E_x$ and $B_y$ 
profiles are on top of each other in vacuum. At time $t = 12000$, the initial pulse does not
exist as it has completely crossed the front-end of the slab. Only the reflected and transmitted pulses exist with their respective 
$E_x$ (blue) and $B_y$ (red) distinctly visible in Fig. \ref{fig:fig2b}. The reflected pulse is propagating along the -$z$-direction, while
the transmitted pulse is inside the dielectric slab and propagating along the $z$-direction. From Faraday's law in \eqref{1.2b}, 
we expect the reflected $E_x$ and $B_y$ to be out of phase. The simulation results in Fig. \ref{fig:fig2b} are in agreement. Furthermore,
from \eqref{2c12}, since $n_1 < n_2$, it is the reflected electric field that will flip sign with respect to the electric field of
the incoming pulse. For the transmitted pulse, from \eqref{1.2c} we note that $B_y / E_x  =  n_2  = 1.5$, and the two field components are in phase. The
simulation results are in accordance with these theoretical expectations. From \eqref{2c12} and \eqref{2c13}, the maximum amplitudes
of the reflected and transmitted pulses are, $\big| E_{r1} / E_0 \big| = 0.2$ and $\big| E_{t1} / E_0 \big| = 0.9798$, respectively. The subscript
$1$ indicates the first encounter with the vacuum-dielectric interface.

\begin{figure}[htp]
\begin{center}
\captionsetup[subfigure]{justification=centering}
\subfigure[\label{fig:fig3a} \ Pulse profiles at $t = 19000$. ]{
       {\includegraphics[width=0.45\textwidth]{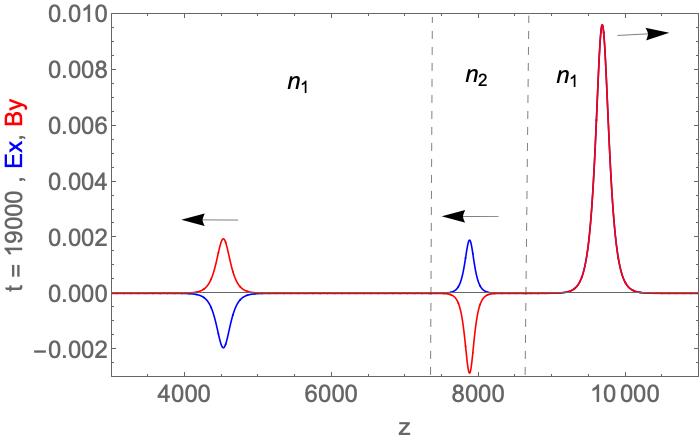}} 
} 
\subfigure[\label{fig:fig3b} \  Pulse profiles at $t = 25000$. ]{
       {\includegraphics[width=0.45\textwidth]{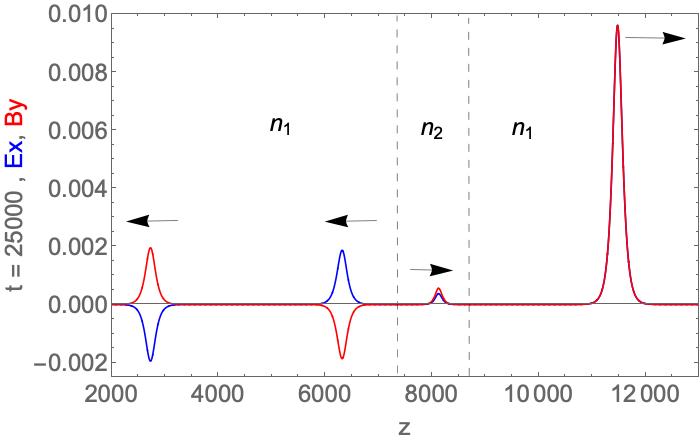}} 
} 
        \captionsetup{justification=raggedright,singlelinecheck=false} 
        \caption{ \label{fig:fig3} (a) Simulation results after scattering from the back-end of the dielectric slab. 
(b) Different pulses after the second scattering event at the front-end of the dielectric slab.  
}
\end{center}
\end{figure}

As the simulation continues to advance in time, the transmitted pulse reaches the back-end of the dielectric slab. Part of this pulse will
get transmitted out into vacuum, and part of it will get reflected and remain inside the dielectric slab. Figure \ref{fig:fig3a} shows this
stage. The pulse transmitted through the back-end of the slab will have $E_x = B_y$ from Faraday's law. From \eqref{2c13}, the transmitted
electric field has the same phase as the wave incident on the back-end. Moreover, the ratio of the electric field of the transmitted
pulse to the incident pulse is $E_{t2} / E_{t1} = 0.9798$. For the pulse reflected from the back-end of the slab, \eqref{2c12} gives
$E_{r2} / E_{t1} = 0.2$ with the two electric fields being in phase. Thus, the reflected and incident electric fields have
the same phase after scattering from the back-end of the dielectric slab. 
In order to satisfy Faraday's law, for the reflected pulse, the magnetic field has to be out of phase with the electric field. 
The results in Fig. \ref{fig:fig3b} are in accordance with these properties of the pulses.

For longer times, the pulse reflected from the back-end, reaches the front-end dielectric boundary and undergoes a reflection and transmission. 
Figure \ref{fig:fig3b} shows the various pulses that are propagating in vacuum and in the dielectric slab.
It is interesting to note that the electric and magnetic fields of the two pulses propagating in vacuum to the 
left of the dielectric slab are out of phase with respect to each other. The phase difference can be explained using
the same arguments, based on Faraday's law and \eqref{2c13}, as discussed above.
Furthermore, for the wave transmitted through the front end, $E_{t3} / E_{r2} = 0.9798$.

From our theory we know that the width of the reflected pulse is the same as the width of the incident pulse. However, the width of
the transmitted pulse either narrows or broadens depending on whether the incident pulse is propagating from a medium of lower
refractive index to a medium of higher refractive index or vice-versa. This is visually discernable in figures \ref{fig:fig2b}, \ref{fig:fig3a},
and \ref{fig:fig3b}. An important conclusion from the analytical analysis is that the maximum amplitude of the transmitted pulse is
different from that obtained for a single plane wave model. The maximum amplitude of the transmitted pulse changes by a factor of
$\sqrt{n_2 / n_1}$ for propagation from a medium with refractive index $n_1$ to a medium with refractive index $n_2$. These results are
borne out as shown in Table \ref{table:t1}. The peak amplitudes of the pulses at various stages of propagation are
in excellent agreement with theory.

\begin{table}[h!]
\label{}
\begin{tabular}{ |>{\centering}p{1.5cm}||>{\centering}p{1.5cm}|>{\centering}p{1.5cm}|>{\centering}p{2cm}|>{\centering}p{2cm}|>
                 {\centering}p{1.8cm}|>{\centering}p{1.8cm}|>{\centering\arraybackslash}p{2cm}|  }
\hline 
               time & $z$   & pulse & $B_{max}$ & $E_{max}$ & $E_t / E_i$ & $E_r / E_i$ & $B_{max} / E_{max}$  \\
$\left( \times 10^{3} \right)$ & & & $\left( \times 10^{-3} \right)$ & $\left( \times 10^{-3} \right)$ & & & \\
\hhline{|=#=|=|=|=|=|=|=|}
\multirow{2}{*}{12} & 6615  & R1V   & 1.9505   & -1.9505  &         & -0.1951  & -1.0    \\ \cmidrule[0.35pt](l){2-8}
                    & 7926  & T1D   & 14.6973  & 9.7982   & 0.9799  &          & 1.5 
                    \\ \midrule[0.75pt] 
\multirow{3}{*}{19} & 4523  & R1V   & 1.9505   & -1.9505  &         &          & -1.0    \\ \cmidrule[0.35pt](l){2-8}
                    & 7876  & R2D   & -2.8556  & 1.9037   &         & 0.1943   & -1.5    \\ \cmidrule[0.35pt](l){2-8}
                    & 9682  & T2V   & 9.6094   & 9.6094   & 0.9807  &          & 1.0 
                    \\ \midrule[0.75pt] 
\multirow{4}{*}{25} & 2730  & R1V   & 1.9504   & -1.9504  &         &          & -1.0    \\ \cmidrule[0.35pt](l){2-8}
                    & 6319  & T1V   & -1.8668  & 1.8668   & 0.9806  &          & -1.0    \\ \cmidrule[0.35pt](l){2-8}
                    & 8124  & R1D   & 0.5563   & 0.3709   &         & 0.1948   & 1.4999  \\ \cmidrule[0.35pt](l){2-8}
                    & 11475 & T2V   & 9.6088   & 9.6088   &         &          & 1.0 
                    \\ 
\hline 
\end{tabular}
\caption{Results from QLA simulations of an initial hyperbolic secant pulse.  
The first column is the simulation time, which correspond to Figs. \ref{fig:fig2b}, \ref{fig:fig3a}, and \ref{fig:fig3b}. 
The second column is the $z$ location of the peak of a pulse in the simulation domain. The third column gives more information 
about the pulse: R and T stand for reflected and transmitted pulses, respectively; $1$ and $2$ indicate the front-end and the back-end, 
respectively, of the dielectric where the pulse is generated; $V$ and $D$ stand, respectively, for vacuum and the dielectric slab 
where the pulse is propagating. $B_{max}$ and $E_{max}$ are the maximum values of $B_y$ and $E_x$, respectively, for a pulse. 
$E_t / E_i$ and $E_r / E_i$ are, respectively, the transmitted and reflected $E_y$ at the peak of a pulse normalized to the 
maximum $E_y$ of the incoming pulse at a given vacuum-dielectric interface. From the analytical model, with 
$\sqrt{n_2 / n_1}$ included in the expression for the transmitted wave amplitude \eqref{2c13}, we get $E_t / E_i = 0.9798$ for all the 
pulses. In addition, from \eqref{2c12}, $\big| E_r / E_i \big| = 0.2$. The results in columns 6 and 7 are in excellent agreement with 
these values. 
}
\label{table:t1}
\end{table}

\subsection*{Poynting flux}
\label{sec:5AA}

\begin{figure}[ht]
    \begin{center}
       {\scalebox{0.3} {\includegraphics{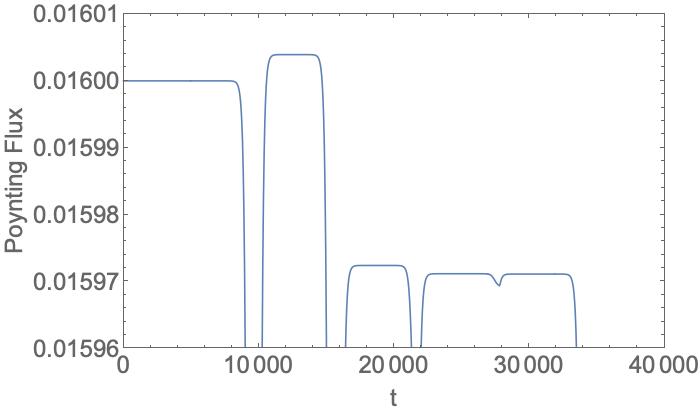}}}
        \captionsetup{justification=raggedright,singlelinecheck=false} 
        \caption{The instantaneous Poynting flux $S(t)$ as a function of time. The presence of the dips is discussed in the text. } 
\label{fig:fig4}
    \end{center} 
\end{figure}

The instantaneous Poynting flux is defined as,
\begin{equation}
S(t) \ = \ \int_0^L dz \ \mathbf{E} \left(z, t \right) \times \mathbf{B} \left(z, t \right) \cdot \mathbf{\hat{n}},
\label{poynt}
\end{equation}
where $\mathbf{\hat{n}}$ is the outward pointing normal at the vacuum-dielectric interface.
In Fig. \ref{fig:fig4}, we plot $S(t)$ as a function of time. The QLA conserves the Poynting flux 
reasonably well over the entire time of the simulation, except for certain gaps in time. In the first gap
around $t = 9000$, the initial pulse is in the vicinity of the front-end boundary of the dielectric
where the incident and reflected pulses overlap and are not clearly separated. The second
gap is around the time when the initial transmitted pulse reaches the back-end boundary of the dielectric.
It is the same with the other gaps when the pulse inside the dielectric reaches one boundary or the other.
In these cases, it is not straightforward to explicitly isolate the incident and reflected pulses, and easily assign
the outward pointing normal to each pulse as required in \eqref{poynt}. In the Appendix, we discuss one possible
method for working around this dilemma. Regardless, it should be noted that $S(t)$ is well conserved away from
these isolated instances of time.

\subsection{Time evolution of an exponential cusp pulse}
\label{sec:5B}

\begin{figure}[htp]
\begin{center}
\captionsetup[subfigure]{justification=centering}
\subfigure[\label{fig:fig5a} \ Initial pulse profile of $E_x$ and $B_y$. ]{
       {\includegraphics[width=0.45\textwidth]{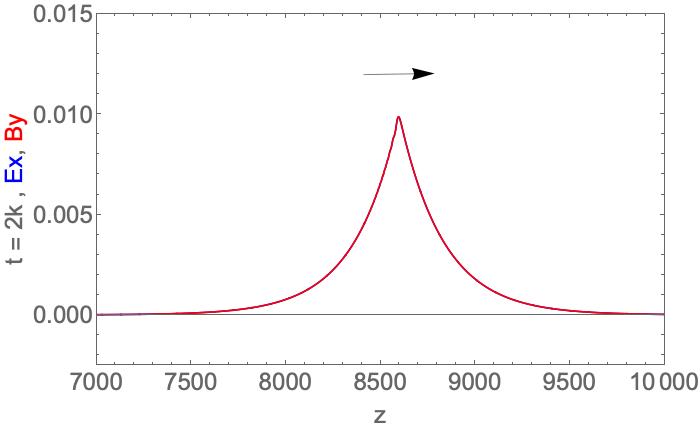}} 
} 
\subfigure[\label{fig:fig5b} \ Pulse profiles at $t = 20000$. ]{
       {\includegraphics[width=0.45\textwidth]{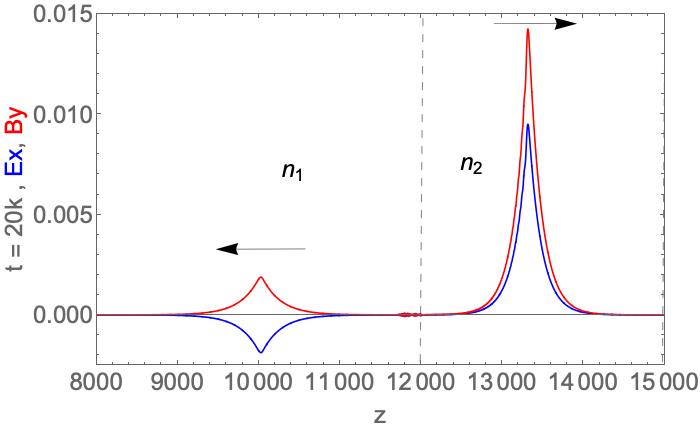}} 
} 
        \captionsetup{justification=raggedright,singlelinecheck=false} 
        \caption{ \label{fig:fig5} The spatial span of the dielectric slab is $12000 \le z \le 15000$. The plots show 
                                   $E_x$ (blue) and $B_y$ (red): (a) the initial exponential cusp pulse in vacuum has $E_x = B_y$; 
                                   (b) reflected and transmitted pulses following the passage of the initial pulse 
                                       through the front-end of the dielectric. 
}
\end{center}
\end{figure}

\begin{figure}[htp]
\begin{center}
\captionsetup[subfigure]{justification=centering}
\subfigure[\label{fig:fig6a} \ Pulse profiles at $t = 38000$. ]{
       {\includegraphics[width=0.45\textwidth]{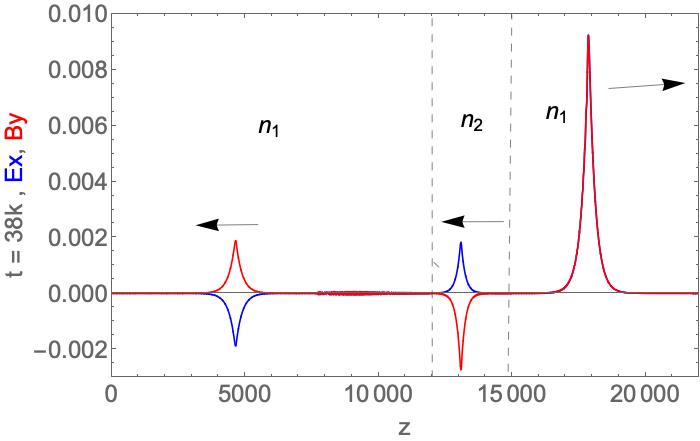}} 
} 
\subfigure[\label{fig:fig6b} \ Pulse profiles at $t = 48500$. ]{
       {\includegraphics[width=0.45\textwidth]{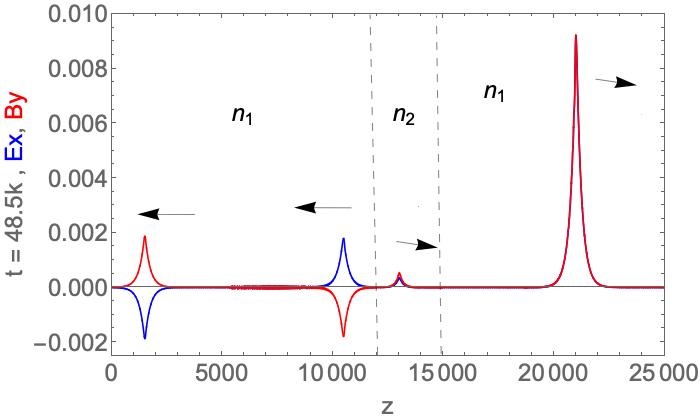}} 
} 
        \captionsetup{justification=raggedright,singlelinecheck=false} 
        \caption{ \label{fig:fig6} (a) Different pulses following scattering from the back-end of the dielectric.   
              (b) The pulses after second scattering from the front-end of the dielectric. 
}
\end{center}
\end{figure}

In this section, we consider the propagation of an exponential cusp profile \eqref{profile2} through a similar dielectric slab
shown in Fig. \ref{fig:fig1a}. Since this pulse has a longer tail than the hyperbolic secant pulse, we increase
the spatial domain for the simulations as well as the spatial extent of the dielectric slab: $12000 \la z  \la 15000$.
The initial profile for the fields is shown in Fig. \ref{fig:fig5a}. The subsequent propagation and splitting of the
initial pulse into reflected and transmitted pulses are shown in Figs. \ref{fig:fig5b}, \ref{fig:fig6a}, and \ref{fig:fig6b}.
The similarity with the propagation of a hyperbolic secant pulse, discussed in the previous subsection, is quite obvious.
Consequently, all analysis of the evolution and splitting of pulses is essentially the same for the two profiles.

The numerical results shown in Table \ref{table:t2} not only agree with the analytical calculations but bear remarkable
resemblance to those in Table \ref{table:t1}. The theoretical formulation was for an electromagnetic Gaussian pulse. However, the QLA
simulations for three different pulse shapes, including a Gaussian, have been in remarkable agreement with the theory. 
Consequently, we believe, that the physics of scattering by a dielectric interface is
common for different pulse shapes that are, initially, spatially confined.

\begin{table}[h!]
\label{}
\begin{tabular}{ |>{\centering}p{1.5cm}||>{\centering}p{1.5cm}|>{\centering}p{1.5cm}|>{\centering}p{2cm}|>{\centering}p{2cm}|>
                 {\centering}p{1.8cm}|>{\centering}p{1.8cm}|>{\centering\arraybackslash}p{2cm}|  }
\hline 
time & $z$ & pulse & $B_{max}$ & $E_{max}$ & $E_t / E_i$ & $E_r / E_i$ & $B_{max} / E_{max}$  \\
$\left( \times 10^{3} \right)$ & & & $\left( \times 10^{-3} \right)$ & $\left( \times 10^{-3} \right)$ & & & \\
\hhline{|=#=|=|=|=|=|=|=|}
\multirow{2}{*}{20}    & 10028 & R1V & 1.8808  & -1.8818 &        & -0.1917 & 0.9995   \\ \cmidrule[0.35pt](l){2-8}
                       & 13317 & T1D & 14.2499 & 9.5     & 0.9677 &         & 1.5 
                       \\ \midrule[0.75pt] 
\multirow{3}{*}{38}    & 4647  & R1V & 1.8773  & -1.8777 &        &        &  0.9998   \\ \cmidrule[0.35pt](l){2-8}
                       & 13087 & R2D & -2.737  & 1.8241  &        & 0.192  & -1.5005   \\ \cmidrule[0.35pt](l){2-8}
                       & 17858 & T2V & 9.2389  & 9.2374  & 0.9724 &        &  1.0001 
                       \\ \midrule[0.75pt] 
\multirow{4}{*}{48.5}  & 1510  & R1V & 1.8756  & -1.8755 &        &        & 1.0       \\ \cmidrule[0.35pt](l){2-8}
                       & 10490 & T1V & -1.7857 & 1.7855  & 0.9788 &        & 1.0001    \\ \cmidrule[0.35pt](l){2-8}
                       & 13015 & R1D & 0.5337  & 0.3558  &        & 0.195  & 1.5       \\ \cmidrule[0.35pt](l){2-8}
                       & 20996 & T2V & 9.2176  & 9.2176  &        &        & 1.0 
                       \\ 
\hline 
\end{tabular}
\caption{Results from QLA simulations of an initial exponential cusp pulse.  
The notation used is the same as in Table \ref{table:t1}. }
\label{table:t2}
\end{table}

\begin{figure}[ht]
    \begin{center}
       {\scalebox{0.3} {\includegraphics{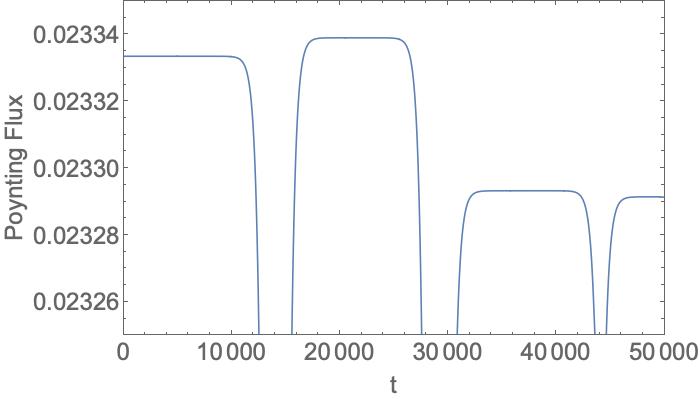}}}
        \captionsetup{justification=raggedright,singlelinecheck=false} 
        \caption{The instantaneous Poynting flux $S(t)$ as a function of time. The presence of the dips is discussed in the text. } 
\label{fig:fig7}
    \end{center} 
\end{figure}

The instantaneous Poynting flux $S(t)$ is plotted as a function of time in Fig. \ref{fig:fig7}. This plot is similar
to that for the hyperbolic secant profile shown in Fig. \ref{fig:fig4}. The explanation put forth for the hyperbolic
secant profile applies here as well.

\section{Summary and Conclusion}

We have shown, analytically and computationally, that the scattering of a spatially confined electromagnetic 
pulse by an interface, separating two disparate dielectric media, is different from the scattering of a plane
wave. In particular, the transmission coefficient is modified by a factor $\sqrt{n_2 / n_1}$ for a pulse
travelling from a medium with refractive index $n_1$ to a medium with a refractive index $n_2$. The analytical
model is based on a Fourier expansion of a Gaussian pulse and the matching of electromagnetic boundary conditions
at the interface separating the two media. The computational results are obtained from a code which
has several layers of formalism embedded into it. We start off by expressing Maxwell equations in a matrix form using
the Reimann-Silberstein-Weber vectors \cite{khan}. This is a 8-spinor representation of Maxwell equations
and has similarities to the Dirac equation for a massless particle. It also forms a basis for the quantum lattice
algorithm that solves Maxwell equations. Since each spinor can be cast as a qubit, the initial
expectation is that we need a 8-qubit algorithm. However, for wave propagation along the $z$-direction, the Pauli spin matrix
$\sigma_z$ is diagonal and does not entangle the qubits. As a result, we developed a 16-qubit algorithm which allows for
entanglement of qubits. The subsequent QLA is a series of streaming and collision operators that advance the 16 qubits from
one lattice site to another and entangles them at each site. The QLA recovers the full set of four Maxwell equations
when expanded to second order in $\epsilon$ -- the separation between adjacent lattice sites. 

The maximum amplitudes of the scattered waves obtained from QLA simulations are in excellent agreement with those
given by the analytical model. The QLA simulations for two different initial pulses lead to the same results for
the maximum amplitudes of the reflected and transmitted pulses. We did not display any results for a Gaussian pulse
of the type used in the theoretical model, since QLA simulations yielded the same ratios for the amplitudes as shown 
in Table \ref{table:t1}. Hence, the theoretical results are applicable to the scattering of
different pulse shapes of finite spatial width.

Finally, we like to point out that the simulation results did not change as the parameter $\epsilon$ was varied. 
We used $\epsilon = 0.2$ for the results displayed in this paper. Even for values of $\epsilon$ approaching unity,
we still recovered the factor $\sqrt{n_2 / n_1}$ associated with the transmission coefficient.

\section{Acknowledgements}

The research was supported by Department of Energy grants DE-SC0021647, DE-FG02-91ER-54109, DE-SC0021651, DE-SC0021857, and
DE-SC0021653.

\appendix*
\section{Determination of the Poynting flux when the pulses overlap}

The dips in the instantaneous Poynting flux in Figs. \ref{fig:fig4} and \ref{fig:fig7} occur when the
incident and reflected pulses overlap in the vacuum-dielectric transition region. 
In such instances, the assignment of an outward pointing normal for the appropriate fields is tricky.
In this section, we resolve this issue for a Gaussian pulse. 

\begin{figure}[ht]
    \begin{center}
       {\scalebox{0.2} {\includegraphics{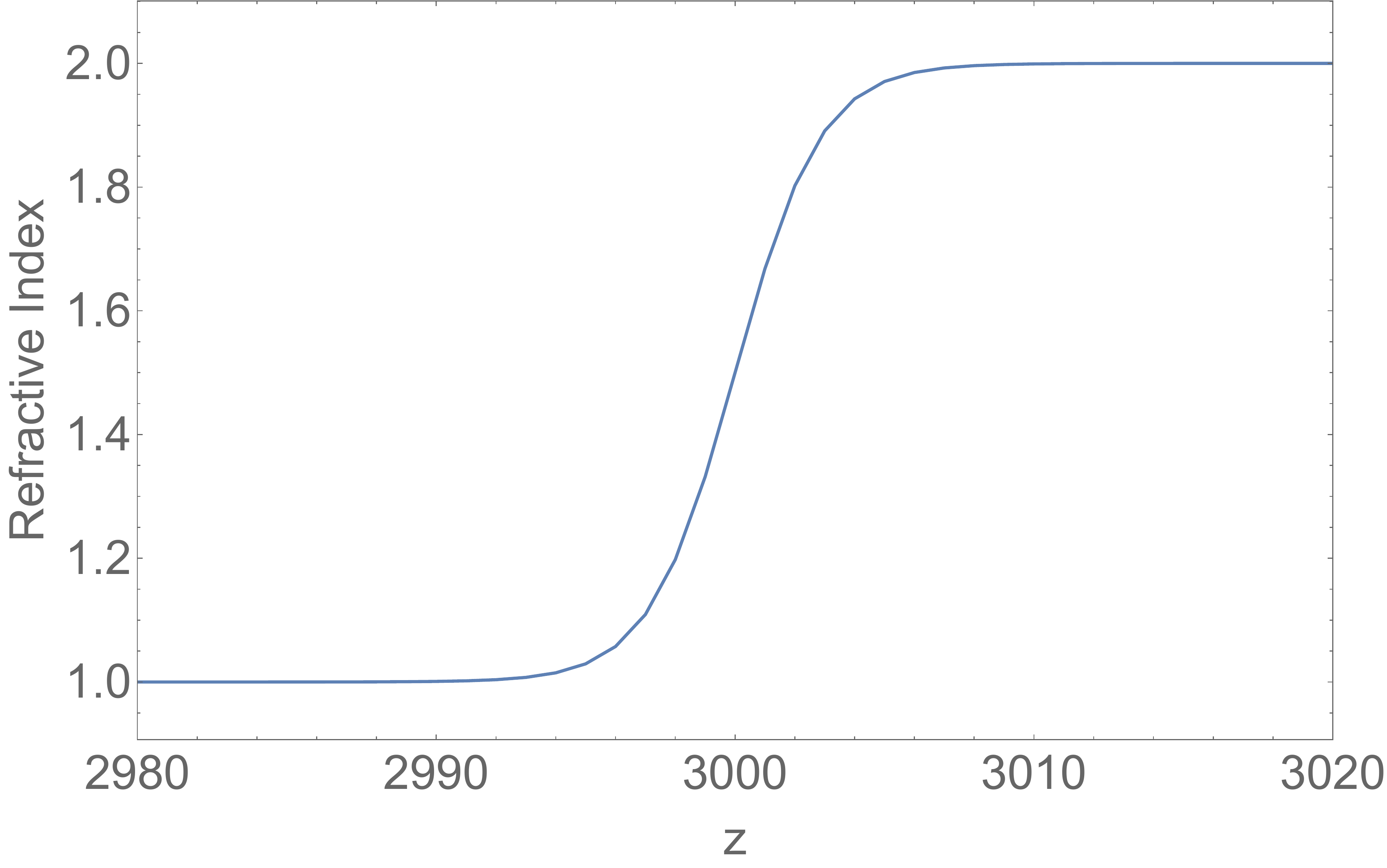}}}
        \captionsetup{justification=raggedright,singlelinecheck=false} 
        \caption{The transition layer, from vacuum to a dielectric with refractive index $n_2 = 2$, is 12 lattice 
                 units wide. 
}
\label{fig:fig8}
    \end{center} 
\end{figure}

\begin{figure}[htp]
\begin{center}
\captionsetup[subfigure]{justification=centering}
\subfigure[\label{fig:fig9a} \ ]{
       {\includegraphics[width=0.45\textwidth]{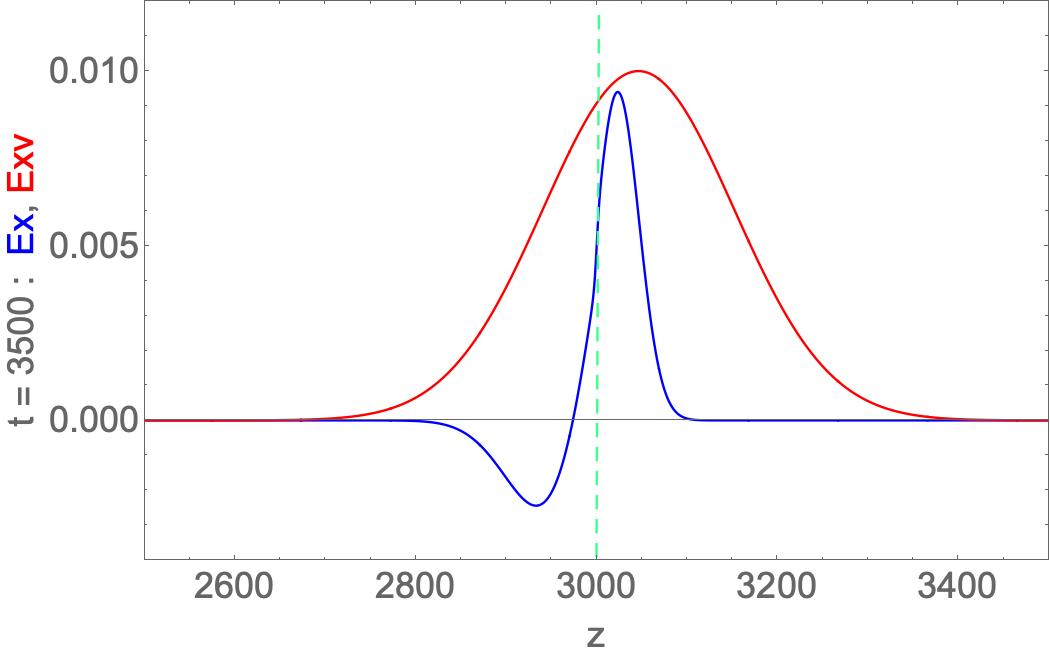}} 
} 
\subfigure[\label{fig:fig9b}]{
       {\includegraphics[width=0.45\textwidth]{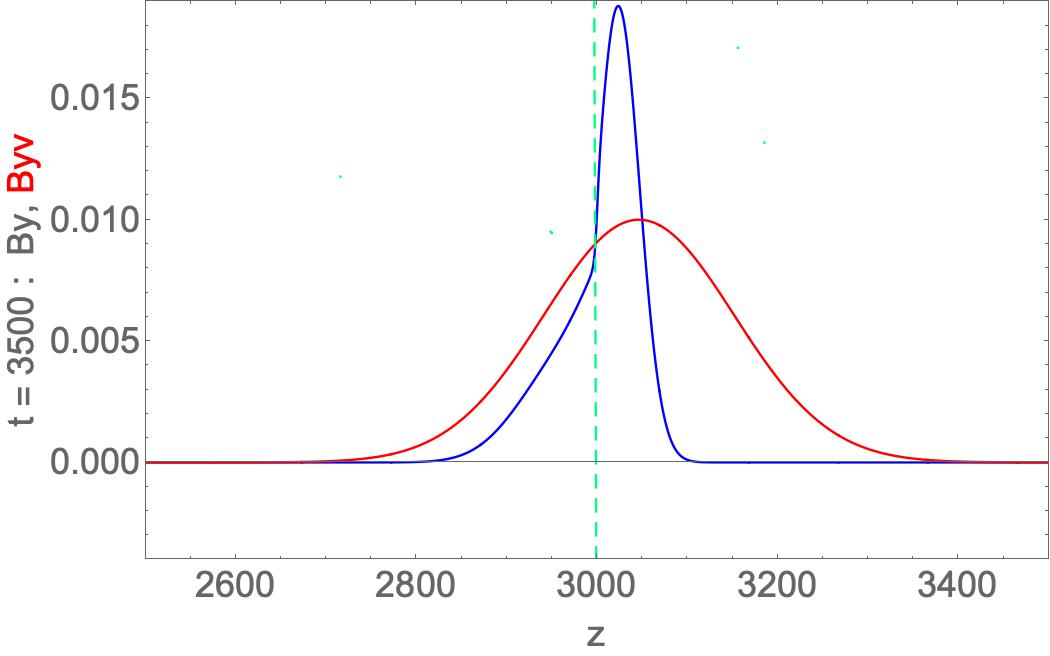}} 
} 
        \captionsetup{justification=raggedright,singlelinecheck=false} 
        \caption{ \label{fig:fig9} (a) The $E_x$ field component at $t = 3500$ (blue). The reversal in $E_x$ is due to  
                                       reflection at the interface. Superimposed in  red is $E_x$ of the initial Gaussian pulse in vacuum.  
                                   (b) The $B_y$ field component at $t = 3500$ (blue). Superimposed in 
                                       red is $B_y$ of the initial Gaussian pulse in vacuum. 
}
\end{center}
\end{figure}

A new set of QLA simulations are performed in which the initial pulse has a Gaussian profile.
The pulse propagates from the vacuum towards a dielectric medium with refractive index $n_2 = 2$.
The vacuum dielectric boundary layer shown in Fig. \ref{fig:fig8} has a transition layer that is
12 lattice units wide. Figures \ref{fig:fig9a} and \ref{fig:fig9b} display, in blue, the electric field $E_x$ and the
magnetic field $B_y$, respectively, of the pulse at $t = 3500$. At this time, 
the incident and reflected pulses overlap in the vicinity of the transition layer. 
We subsequently carry out QLA simulations in vacuum and superimpose the associated $E_x$ and $B_y$ fields, in red, 
in Figs. \ref{fig:fig9a} and \ref{fig:fig9b}, respectively.

\begin{figure}[ht]
    \begin{center}
       {\scalebox{0.25} {\includegraphics{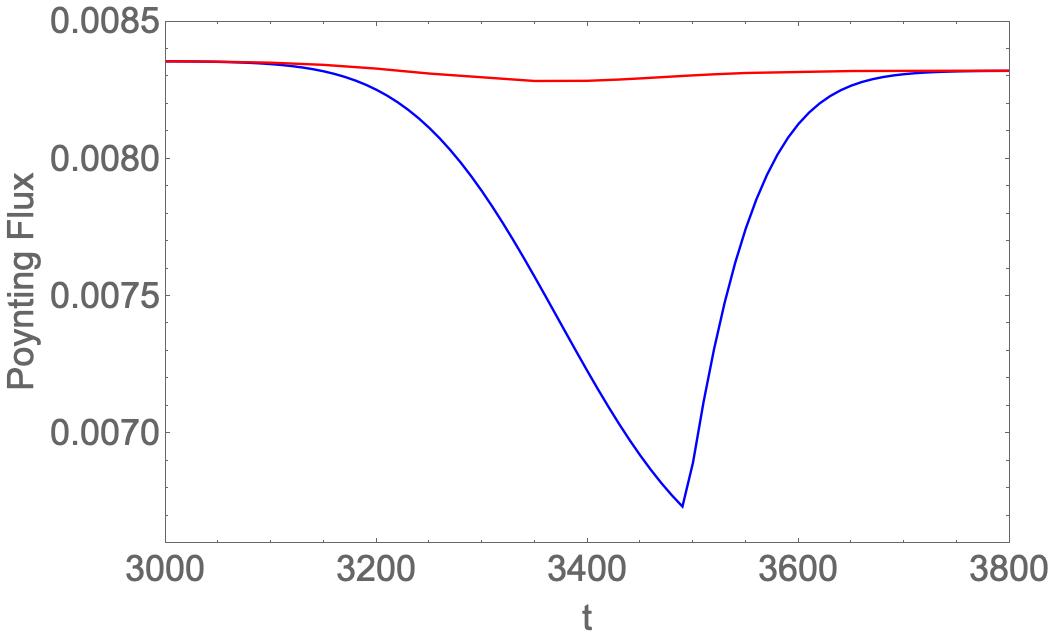}}}
        \captionsetup{justification=raggedright,singlelinecheck=false} 
        \caption{In red is the instantaneous Poynting flux as a function of time evaluated using the subtraction technique. In
                 blue is the result obtained using the same procedure that led to gaps in Figs. \ref{fig:fig4} and \ref{fig:fig7}.
}
\label{fig:fig10}
    \end{center} 
\end{figure}

In order to correctly determine the reflected part of the pulse and associate it properly with the
outward pointing normal, we subtract, in Figs. \ref{fig:fig9a} and \ref{fig:fig9b}, 
the fields in blue from the vacuum fields in red for $z < 3000$. We know that for $z > 3000$ there is
only the transmitted pulse, while for $z < 3000$ the initial and the reflected pulses coexist.
The subtraction separates out the reflected pulse from the incident pulse; the incident pulse propagating
along the $z$-direction while the reflected pulse propagates along the -$z$-direction. This procedure helps remove any
ambiguity, and allows for correct evaluation of $S(t)$. It has to be carried out for all the time steps during which
the incident and reflected pulses overlap. In Fig. \ref{fig:fig10}, we compare the results from this procedure
with the one used to evaluate $S(t)$ in Figs. \ref{fig:fig4} and \ref{fig:fig7}. 
Figure \ref{fig:fig10} illustrates that the modified evaluation of $S(t)$ conserves the flow of energy when we properly
account for the reflected part of the pulse in the overlap region. The dips in Figs. \ref{fig:fig4} and \ref{fig:fig7}
are just due to the approximate way in which we try to separate the incident and reflected pulses.

Earlier in this paper it was mentioned that, through QLA simulations, we can visualize the interplay between incident, 
reflected, and transmitted pulses in the neighborhood of the transition region between two dielectric media. 
The blue curves in Figs. \ref{fig:fig9a} and \ref{fig:fig9b} illustrate that point. This would be difficult to realize
with the theoretical model since it does not solve an initial value problem.

\end{document}